\documentclass[prb,num, preprint, showpacs, 10pt, twocolumn,superscriptaddress]{revtex4-1}
\usepackage{graphicx}
\usepackage[utf8]{inputenc}
\usepackage[colorlinks=true, pdfstartview=FitV, linkcolor=blue, citecolor=blue]{hyperref}
\usepackage{color}
\usepackage{colortbl}
\usepackage{epstopdf}
\usepackage{graphicx}
\usepackage{amsmath}
\usepackage{array,multirow}
\usepackage[table]{xcolor}% http://ctan.org/pkg/xcolor
\usepackage{tabularx}
\usepackage{float}
\usepackage{ctable}
\makeatletter
\def\hlinewd#1{%
\noalign{\ifnum0=`}\fi\hrule \@height #1 %
\futurelet\reserved@a\@xhline}

\begin{document}

\preprint{}

\title{$\beta$-As$_2$Te$_3$: Pressure-Induced 3D Dirac Semi-Metal with Room-Pressure Ultra-Low Lattice Thermal Conductivity}

\author{E. Lora da Silva}
\email{estelina.silva@fc.up.pt}
\affiliation{IFIMUP, Departamento de F\'{i}sica e Astronomia, Faculdade de Ci\^{e}ncias da Universidade do Porto, Porto, Portugal}
\affiliation{Instituto de Dise\~{n}o para la Fabricaci\'{o}n y Producci\'{o}n Automatizada, MALTA Consolider Team, Universitat Polit\`{e}cnica de Val\`{e}ncia, Val\`{e}ncia, Spain}

\author{A. Leonardo}
\affiliation{Departmento de F\'{i}sica, MALTA Consolider Team, Universidad del País Vasco, UPV/EHU, Spain}
\affiliation{Donostia International Physics Center (DIPC), Donostia, Spain}

\author{Tao Yang}
\affiliation{College of New Materials and New Energies, Shenzhen Technology University, Shenzhen, 518118, China}

\author{M. C. Santos}
\affiliation{Instituto de Dise\~{n}o para la Fabricaci\'{o}n y Producci\'{o}n Automatizada, MALTA Consolider Team, Universitat Polit\`{e}cnica de Val\`{e}ncia, Val\`{e}ncia, Spain}

\author{R. Vilaplana}
\affiliation{Centro de Tecnologías F\'{i}sicas: Ac\'{u}stica, Materiales y Astrof\'{i}sica, MALTA Consolider Team, Universitat Polit\`{e}cnica de Val\`{e}ncia, Val\`{e}ncia, Spain}

\author{S. Gallego-Parra}
\affiliation{Instituto de Dise\~{n}o para la Fabricaci\'{o}n y Producci\'{o}n Automatizada, MALTA Consolider Team, Universitat Polit\`{e}cnica de Val\`{e}ncia, Val\`{e}ncia, Spain}

\author{A. Bergara}
\affiliation{Departmento de F\'{i}sica, MALTA Consolider Team, Universidad del País Vasco, UPV/EHU, Spain}
\affiliation{Donostia International Physics Center (DIPC), Donostia, Spain}
\affiliation{Centro de F\'{i}sica de Materiales CFM, Centro Mixto CSIC-UPV/EHU, Donostia, Spain}

\author{F. J. Manj\'{o}n}
\affiliation{Instituto de Dise\~{n}o para la Fabricaci\'{o}n y Producci\'{o}n Automatizada, MALTA Consolider Team, Universitat Polit\`{e}cnica de Val\`{e}ncia, Val\`{e}ncia, Spain}

\begin{abstract}
An \textit{ab-initio} study of beta-As2Te3 ($R\bar{3}m$ symmetry) at hydrostatic pressures shows that this compound is a trivial small band-gap semiconductor at room pressure that undergoes a quantum topological phase transition to a 3D topological Dirac semi-metal around 2 GPa. At higher pressures, the band-gap reopens and again decreases above 4 GPa. Our calculations predict an insulator-metal transition above 6 GPa due to the closing of the band-gap, with strong topological features persisting between 2 and 10 GPa with Z$_4$=3 topological index. By investigating the lattice thermal-conductivity ($\kappa_\textrm{L}$), we observe that close to room conditions $\kappa_\textrm{L}$ is very low, either for the in-plane and the out-of-plane axis, with 0.098 and 0.023 Wm$^{-1}$K$^{-1}$, respectively. This effect occurs due to the presence of two low-frequency optical modes, namely E$_u$ and E$_g$, which increase the phonon-phonon scattering rate. Therefore, our work suggests that ultra-low lattice thermal-conductivities, which enable highly efficient thermoelectric materials, can be engineered in systems that are close to a structural instability derived from phonon Kohn anomalies. At higher pressures, the values of the in- and out-of-plane thermal-conductivities not only increase in magnitude, but also approximate in value as the layered character of the compound decreases. 
\end{abstract}

%\pacs{}% insert suggested PACS numbers in braces on next line

\maketitle %\maketitle must follow title, authors, abstract and \pacs

\date{\today}

%%%%%%%%%%%%%%%%%%%%%%%%%%%%%%%%%%%%
\section{Introduction}
%%%%%%%%%%%%%%%%%%%%%%%%%%%%%%%%%%%%

Sesquichalcogenides A$_2$X$_3$ (X=S, Se, Te) of group-15 cations (A=As, Sb, Bi) with tetradymite ($R\bar{3}m$) symmetry have stimulated enormous research activity, because of their exceptional thermoelectric properties. Moreover, these systems have attracted even more attention due to their unique fundamental properties since Bi$_2$Se$_3$, Bi$_2$Te$_3$ and Sb$_2$Te$_3$ were discovered as prospective 3D topological insulators (TIs) evidencing a single Dirac cone on the surface.\cite{Science.325.178.2009,NatPhys.5.438.2009} This type of compounds represents a new class of matter with insulating bulk electronic states and topologically protected metallic surface states due to time-reversal symmetry and strong spin-orbit interaction; and with major applications to spintronics and quantum computation.\cite{RevModPhys.82.3045.2010}

Most of the studies performed on the tetradymite-like A$_2$X$_3$ sesquichalcogenides have been applied to compounds with Sb and Bi species and much less attention has been paid to the As-based compounds. These latter systems do not tend to crystallize into the tetradymite structural phase at ambient conditions due to the strong lattice distortions caused by the stereo-active lone electron pair (LEP) of the As cation. In particular, at room conditions As$_2$Te$_3$ crystallizes in the monoclinic \textit{C2/m} phase ($\alpha$-As$_2$Te$_3$), showing some interesting properties and applications, including efficient thermoelectric properties,\cite{JPhysChemSolids.2.181.1957} electrical threshold and memory switching properties for phase change memory (PCM) devices (similar to other group-14 and group-15 chalcogenides).\cite{JNonCrystSolids.24.365.1977,SciRep.9.19251.2019,PhysStatSolRRL.13.1900320.2019} More recently incredible mechanical properties of As$_2$Te$_3$ have been observed with potential applications for superstretchable membranes.\cite{ACSNano.13.10845.2019,JMatChemC.8.2400.2020} 

As$_2$Te$_3$ is also very interesting since it has the ability to display rich polymorphism for different experimental conditions (temperature and/or pressure), such as the metastable tetradymite structure ($\beta$-As$_2$Te$_3$) \cite{JNonCrystSolids.24.365.1977,SSSR.10.1431.1974,JSolidStateChem.74.277.1988,ThermochimActa.186.247.1991} and the low-temperature phase with $P2_1/m$ symmetry ($\beta$'-As$_2$Te$_3$).\cite{InorChem.54.9936.2015} In particular, several theoretical studies have addressed the structural, mechanical and electronic properties of $\beta$-As$_2$Te$_3$ at room conditions \cite{SCHEIDEMANTEL2003667,DENG2016695,PhysRevB.93.245307, doi:10.1002/adfm.201904862, doi:10.1002/adma.201904316} and it has been experimentally demonstrated that this phase also displays good thermoelectric properties.\cite{doi:10.1002/aelm.201400008,doi:10.1063/1.4950947,C8CP00431E} Moreover, recent calculations \cite{doi:10.1002/adfm.201904862,doi:10.1002/adma.201904316} show that $\beta$-As$_2$Te$_3$ exhibits other outstanding properties, for instance as PCM, which are also related to respective TI features. Such properties are the result of an unconventional bonding mechanism known as “metavalent bonding”.\cite{doi:10.1002/adfm.201904862,doi:10.1002/adma.201904316,doi:10.1002/adma.201803777,doi:10.1002/adma.201806280} Finally, another major interest in $\beta$-As$_2$Te$_3$ is the possibility of finding a pressure-induced electronic topological transition (ETT) \cite{doi:10.1002/pssb.201200672}, as in other tetradymite-like sesquichalcogenides, which can result in a significant enhancement of respective thermoelectric properties. \cite{doi:10.1021/cm000888q,PhysRevB.68.085201,doi:10.1063/1.2973201,doi:10.1021/cm902000x}

Very few high-pressure (HP) studies have been devoted to understand the properties of As$_2$Te$_3$. \cite{doi:10.1021/acs.jpcc.6b06049,doi:10.1021/acs.inorgchem.6b00073,ZHANG2016551,SCHEIDEMANTEL20051744} However these show that the $\alpha$-As$_2$Te$_3$ may undergo a trivial semiconductor-metal transition above 4 GPa followed by a phase transitioning directly to a monoclinic structure (phase $\gamma$ S.G. \textit{C2/c}) above 13 GPa; and not to the $\beta$ phase, as it had previously been observed by applying uniaxial stress.\cite{SCHEIDEMANTEL20051744} Moreover, several isostructural phase transitions (IPTs) have also been suggested.\cite{doi:10.1021/acs.inorgchem.6b00073,ZHANG2016551} As regards to the $\beta$-As$_2$Te$_3$ phase, the magnitude of the spin-orbit coupling of As is lower than that of A$_2$Te$_3$ (A=Sb, Bi) due to the lighter mass of As. The transition from a trivial band insulator to a TI would thus require HP application on the $\beta$-As$_2$Te$_3$ system. According to a theoretical study,\cite{doi:10.1063/1.4892941} it is demonstrated that uniaxial strain could cause a quantum topological phase transition (QTPT) from a band insulator to a TI state at 1.8 GPa. Another theoretical study has shown that application of isotropic strain,\cite{doi:10.1063/1.4950947} enables an overlap of the electronic bands at the Fermi level ($\Delta V/V$ $\sim$ -7\%), accompanied by a metallic state transition, characteristic of an ETT.\cite{doi:10.1063/1.4950947} 

Finally, it must be stressed that from first-principles calculations performed on the rhombohedral tetradymite structure of Sb$_2$Se$_3$ (a polymorph of Sb$_2$Se$_3$ that has not yet been experimentally reported), have shown that this compound should evidence a pressure-induced QTPT,\cite{PhysRevB.89.035101} thus transitioning from a trivial semiconductor to a 3D topological Dirac semi-metal (3D TDS).\cite{PhysRevB.89.035101} 

In this work, we report a first-principles study of the electronic and vibrational properties of $\beta$-As$_2$Te$_3$ under hydrostatic compression, ranging between 0 and 12 GPa. We show that this compound undergoes a pressure-induced QTPT around 2 GPa, in which a linear-type dispersion is observed at the $\Gamma$-point, evidencing a transition from a trivial insulator to a 3D Dirac semi-metal. This feature leads to a TI behavior around 2 GPa that persists up to 12 GPa. We also compute the lattice thermal-conductivity at three different pressure ranges, in order to study the influence of hydrostatic pressure to $\kappa_\textrm{L}$ and compare the results of the obtained low values to the phonon lifetimes and low-frequency optical phonon modes.

%%%%%%%%%%%%%%%%%%%%%%%%%%%%%%%%%%%%
\section{Results and Discussion}
%%%%%%%%%%%%%%%%%%%%%%%%%%%%%%%%%%%%
\begin{table}[!]
\begin{center}
\caption{ \label{table:lat_param} Equilibrium lattice parameters.}
\begin{tabular}{| c| c|c|c|}\hlinewd{1pt}
  References &	a$_0$(\AA) & c$_0$(\AA) & V$_0$(\AA$^3$)\\\hline
present (PBEsol) & 4.096	& 30.592  & 444.46\\ \hline
GGA \cite{JSolidStateChem.74.277.1988,Morin2015,Jain2013} & 4.089	& 30.297 & 438.76\\ \hline
optB88-vdW \cite{CHOUDHARY2019300,JARVIS} & 4.075	& 30.306 & 435.79\\ \hline
PBE \cite{C6RA01770C} & 4.102 & 29.745 & 433.40\\ \hline 
Experiment 1 \cite{Morin2015}& 4.047  & 29.498 & 418.40\\\hline
Experiment 2 \cite{doi:10.1002/aelm.201400008} & 4.047  & 29.502 & 418.51 \\\hline
\hlinewd{1pt}
\end{tabular}
\end{center}
\end{table}

\begin{figure}[h]
\begin{center}
\includegraphics[width=5cm]{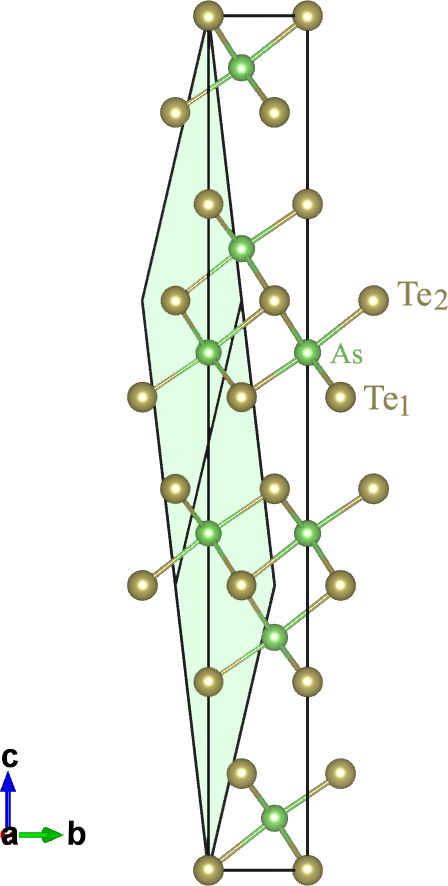}
\caption{\label{fig:structure}
Crystal structure of $\beta$-As$_2$Te$_3$ in the hexagonal unit-cell representation (the rhombohedral primitive-cell is represented in shaded green).} 
\end{center}
\end{figure}

The obtained relaxed lattice parameters for the PBEsol+SoC calculations are shown in Table \ref{table:lat_param} together with other calculations from the literature as well as experimental values. Our results are consistent with the rest of the calculations and overestimate the value of $c_0$ by $3.58\%$ resulting in a larger unit-cell volume when compared to experiments. For visualization purposes the unit-cell is shown in (Fig. \ref{fig:structure}).

 Similar overestimation of lattice parameters were found with PBEsol+SoC calculations for isostructural Sb$_2$Te$_3$. \cite{LAWAL20172302}

%%%%%%%%%%%%%%%%%%%%%%%%%%%%%%%%%%%%
\subsection{Electronic Band Structure as a function of Pressure}
%%%%%%%%%%%%%%%%%%%%%%%%%%%%%%%%%%%%

\begin{figure*}
\begin{center}
\includegraphics[width=18cm]{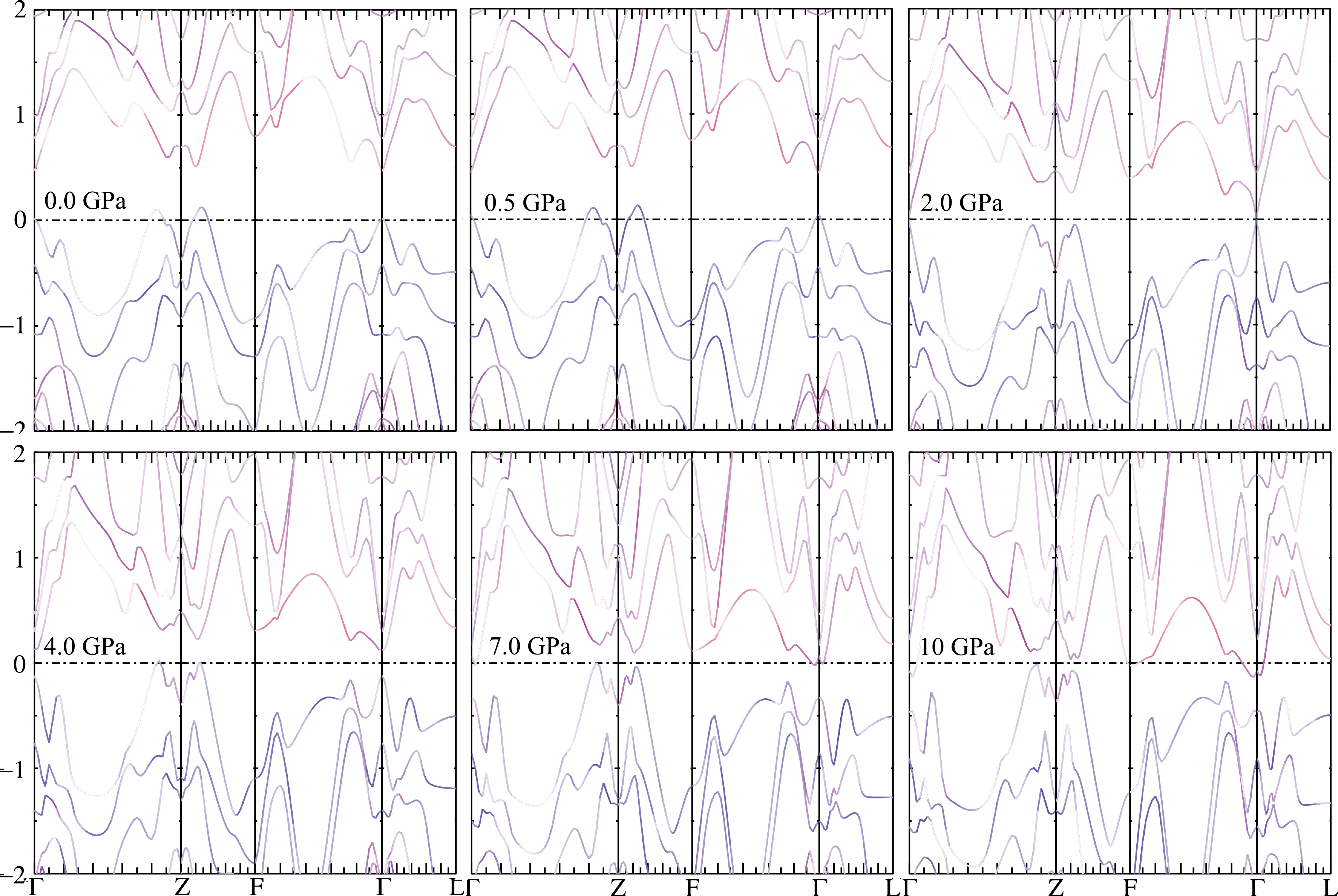}
\caption{\label{fig:band_weight}
The \textit{QS}GW+SoC weighted electronic band structure of $\beta$-As$_2$Te$_3$ for different pressure values. The red dispersion represent the As-\textit{p} states and the blue dispersion the Te-\textit{p} states.} 
\end{center}
\end{figure*}

\begin{figure}[!]
\begin{center}
\includegraphics[width=8cm]{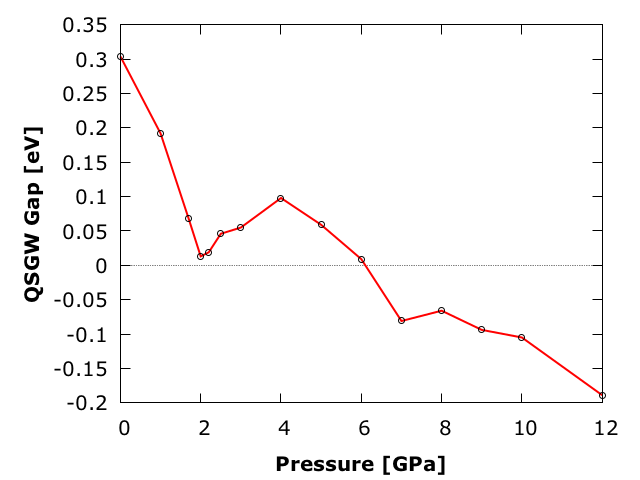}
\caption{\label{fig:gap_pressure}
The \textit{QS}GW+SoC band-gap as a function of pressure of $\beta$-As$_2$Te$_3$.} 
\end{center}
\end{figure}

We have calculated the electronic band dispersions of $\beta$-As$_2$Te$_3$ (represented in $R\bar{3}m$ reciprocal space-group), by employing the quasiparticle self-consistent GW (\textit{QS}GW).\cite{PhysRevLett.96.226402.2006,PhysRevB.76.165106.2007} These were performed for different pressures values, ranging between 0 GPa and 10 GPa (Tab. \ref{table:qsgw_gap} and Figs. \ref{fig:band_weight} and \ref{fig:gap_pressure}). The structures for different pressures were previously minimized using VASP \cite{PhysRevB.47.558.1993,PhysRevB.54.11169.1996} with PBEsol+SoC, which serve as input for the \textit{QS}GW calculations.

At 0 GPa (Fig. \ref{fig:band_weight}) we observe that the conduction band minimum (CBM) and the valence band maximum (VBM) are positioned at different high-symmetry points along the \textbf{Z}-\textbf{F} segment, that we will hereafter denote as the Z' points. As a consequence, $\beta$-As$_2$Te$_3$ exhibits an indirect \textbf{Z}'-\textbf{Z}' band-gap of 0.30 eV, thus making $\beta$-As$_2$Te$_3$ a small band-gap semiconductor at room conditions. Our \textit{QS}GW direct band-gap energy at $\Gamma$ is estimated to be around 0.45 eV at 0 GPa. The present results differ from previous DFT@PBE calculations,\cite{C6RA01770C,Morin2015,DENG2016695} where a direct gap was observed to be between 0.12 and 0.30 eV. Curiously enough, Pal and Waghmare\cite{doi:10.1063/1.4892941} showed that at vanishing strain, the VBM and the CBM are located along different directions of the Brillouin-zone (BZ) evidencing an indirect band-gap around 0.22 eV; with direct band-gap of around 0.35 eV. These latter calculations have been performed by employing an all-electron full potential linearized augmented plane wave (FP-LAPW) technique, with the PBE functional, and by considering SoC effects. Moreover, an indirect band-gap has also been observed for monolayer $\beta$-As$_2$Te$_3$ with a value of 0.72 (1.05) eV, and obtained by employing PBE (HSE) functionals.\cite{PhysRevB.93.245307}

Following with the review of reported values for the band-gap, Scheidemantel \cite{SCHEIDEMANTEL2003667} has found a direct band-gap for $\beta$-As$_2$Te$_3$ with 0.12 eV, where FP-LAPW basis-sets and the PBE functional were applied; while Sharma \cite{SHARMA2011899} found an indirect band-gap along \textbf{Z}-\textbf{F} direction of 0.22 eV by employing similar methodologies.

The differences observed in the character and width of the band-gaps are mainly dependent and sensitive to the applied methodologies. Within this context, we must note that by applying one-shot GW calculations the band-gap character of Sb$_2$Te$_3$ has also shown different conclusions. While Lawal \textit{et al.}\cite{LAWAL20172302} predicted a direct band-gap at $\Gamma$, Nechaev and co-workers\cite{PhysRevB.91.245123} have observed an indirect band-gap along \textbf{Z}'-$\Gamma$.

As pressure increases, the present calculations show variations not only of the electronic band dispersions, but also of the band-gap character and width (see Tab. \ref{table:qsgw_gap} and Fig. \ref{fig:band_weight}). On increasing pressures, the band-gap decreases and the CBM moves towards the $\Gamma$-point. At around 1 GPa the band-gap is indirect along \textbf{Z}'-$\Gamma$ with a value of 0.19 eV. Therefore, at 1 GPa the indirect character of the band-gap (\textbf{Z}'-$\Gamma$) of $\beta$-As$_3$Te$_2$ is similar to that found for Sb$_2$Te$_3$ at 0 GPa.\cite{PhysRevB.91.245123} 

Around 1.7 GPa the band-gap character changes from indirect to direct and is positioned at the $\Gamma$-point. An interesting feature occurs around 2 GPa, where we observe that the direct band-gap at the $\Gamma$-point closes forming a linear-type dispersion. 
This feature is consistent with a previous study that found that the application of an uniaxial strain in the \textbf{Z}-direction of 1.77 GPa induces the system to pass through a Weyl metallic state with a single Dirac cone in its electronic structure at the $\Gamma$-point.\cite{AppPhysLett.105.062105.2014} Our calculations therefore evidence a pressure-induced QTPT from a semiconductor-to-semi-metal, making $\beta$-As$_2$Te$_3$ a three dimensional topological Dirac semi-metal (3D TDS) under hydrostatic pressure. Dirac semi-metals are 3D phases of matter with gapless electronic excitations and are protected by topology and symmetry, with well-defined 3D massless charge carriers. As 3D analogs of graphene, these systems have generated much recent interest. 
These results can be compared to those found for Cd$_3$As$_2$, \cite{PhysRevResearch.1.033101} a system which has attracted intensive research interest as an archetypical TDS that hosts 3D linear-dispersive electronic bands close to the Fermi level at room conditions.\cite{PhysRevMaterials.2.120302} Other intrinsic TDSs are found among the following systems: Bi$_{1-x}$Sb$_x$,\cite{PhysRevLett.111.246603} Cd$_3$As$_2$ \cite{PhysRevMaterials.2.120302} and Na$_3$Bi.\cite{Liu864}

Apart from the QTPT observed close to 2 GPa, compressed $\beta$-As$_2$Te$_3$ shows considerable changes of the VBM and CBM which could be seen as pressure-induced ETTs. An ETT (or Lifshitz transition \cite{SovPhysJETP.11.1130.1960}) occurs when an extreme of the electronic-band structure, which is associated to a Van Hove singularity of the density of states, crosses the Fermi level.\cite{doi:10.1002/pssb.201200672}  

When the pressure over the system is increased beyond the 2 GPa value, calculations show that the gap reopens and it gradually transforms respective character from direct to indirect. The VBM is positioned along the $\Gamma$-\textbf{Z} segment, which we will define as the \textbf{Z}"-point. Above 4 GPa, the band-gap starts to decrease again as the CBM moves slightly away from the $\Gamma$-point (Fig. \ref{fig:gap_pressure}). Beyond 6 GPa, the band-gap closes thus evidencing an insulator-metal transition. The metallic character of the compound persists up to 12 GPa. Such results are consistent with discussion of Refs. \onlinecite{C6RA01770C} and \onlinecite{AppPhysLett.105.062105.2014}. Both report a band inversion with parity reversal of states close to the Fermi level, either by employing uniaxial compression ($\Delta$ V/V $\sim$ -7\%) or isotropic stress ($\Delta$ V/V $\sim$ -5\%), respectively. From the present calculations, hydrostatic pressure at $\sim$ 7 GPa would correspond to a higher compression with $\Delta$ V/V $\sim$ -14.64\%. This feature is consistent with the value of the energy band-gap obtained from the \textit{QS}GW calculations when compared to band-gaps obtained when applying DFT with (semi-)local functionals that are known to underestimate respective magnitudes.

\begin{table}[!]
\begin{center}
\caption{ \label{table:qsgw_gap} \textit{QS}GW+SoC electronic band-gap character for different pressure values.}
\begin{tabular}{| c| c|}\hlinewd{1pt}
  Pressure [GPa] &		Character (VBM-CBM)\\\hline
0.0 &   Indirect (Z'-Z') \\ \hline
1.0 &  Indirect (Z'-$\Gamma$) \\ \hline
1.7 &  Direct ($\Gamma$-$\Gamma$) \\ \hline
2.0 &  Direct ($\Gamma$-$\Gamma$) \\ \hline
2.2 &  Direct ($\Gamma$-$\Gamma$) \\ \hline
2.5 &  Indirect (Z"-$\Gamma$) \\ \hline
3.0 &  Indirect (Z"-$\Gamma$) \\\hline
4.0 &  Indirect (Z"-$\Gamma$)\\\hline
5.0 &  Indirect (Z"-$\Gamma$')\\\hline
6.0 &  Indirect (Z"-$\Gamma$’)\\\hline
7.0 &  Indirect (Z"-$\Gamma$’)\\\hline
8.0 &  Indirect (Z"-$\Gamma$’)\\\hline
9.0 &  Indirect (Z"-$\Gamma$’)\\\hline
10.0 &  Indirect (Z"-$\Gamma$’)\\\hline
12.0 &  Indirect (Z"-$\Gamma$’)\\\hline
\hlinewd{1pt}
\end{tabular}
\end{center}
\end{table}

Fig \ref{fig:band_weight} shows the contributions of the As-\textit{p} and Te-\textit{p} states to the dispersion curves. At 0 GPa, the valence bands are mostly dominated by the \textit{p}-states of Te atoms (blue dispersion curves), whereas the conduction bands are mostly contributed by As-\textit{p} states (red dispersion curves). By further increasing the pressure to values close to the QTPT ($\sim$ 2.0 GPa) we observe that some As states start appearing at the high-symmetry \textbf{Z}-point, which are more evident at 4 GPa, when the band-inversion is observed. Moreover, at around the QTPT a fraction of As states can be observed at the VBM, at the $\Gamma$-point. At 4 GPa, when the band-gap reopens we observe that the small fraction of the As-states persists at the $\Gamma$-point.

In order to confirm the possibility of topological invariants for the different pressure values, we have performed a topological analysis of the eigenvalues at the high-symmetry \textbf{K}-points by employing the Check Topological Materials Tools as detailed in Ref. \onlinecite{Nature.566.480.2019}. We then obtain a set of irreducible representations at each maximal \textbf{K}-vec. By using the compatibility relations and the set of Elementary Band Representations (EBRs), it is possible to probe whether the set of bands can be linearly combined as EBRs (Tab. \ref{table:topo} and in Sup. material Tab. \ref{table:topo_list}). 

The topology analysis performed on our system determines that in the 0-2 GPa range we have a trivial insulator; while for pressures above 2 GPa $\beta$-As$_2$Te$_3$ the compound is a TI that belongs to a strong topological class with Z$_4$=3 topological index (Tab. \ref{table:topo}). These systems are known as Split Elementary Band Representations (SEBR). The bands directly below and above the Fermi level form a EBR, with a topological band-gap.\cite{Nature.566.480.2019} SEBRs can be tuned either to be topologically nontrivial insulators or to be semi-metals. 

The characteristic electronic properties of topological features observed around 2 GPa (see Fig. \ref{fig:band_weight}), which corresponds to $\Delta$ V/V $\sim$ -7.30\%, lead to protected surface states and novel responses to applied electric and magnetic fields.\cite{RevModPhys.90.015001}
The $\beta$-As$_2$Te$_3$ system therefore can be compared to prototypical graphene with large spin-orbit coupling where the VBM and CBM touch at the Fermi level.  

Since the effect of pressure can be mimicked by the effects of chemical doping, our present results opens the possibility of obtaining a 3D TDS close to room conditions when introducing substitutional impurities to $\beta$-As$_2$Te$_3$, namely Sb or Bi.

It is noteworthy of mentioning that the Dirac cone which occurs in $\beta$-As$_2$Te$_3$, close to 2 GPa, is only observed at the $\Gamma$-point, unlike other 3D TDSs, e.g. Cd$_3$As$_2$, Na$_3$Bi, ZrTi$_5$ and bP.\cite{PhysRevB.93.195434} Therefore, Kohn anomalies, associated to TDS, can only occur at the BZ center. This feature would inhibit any possibility of occurring a Fermi-surface nesting and thus any Peierls distortion in tetradymite-like As$_2$Te$_3$, and likely in any other isostructural group-15 sesquichalcogenides. Our results contradict a previous study that suggested that Peierls distortion may occur for group-15 sesquichalcogenides,\cite{doi:10.1002/adma.201908302} however confirms the previous questioning of the occurrence of this distortion in these type of materials.\cite{doi:10.1021/acs.inorgchem.0c01086}

\begingroup
\squeezetable
\begin{table}[!]
\begin{center}
\caption{ \label{table:topo} }
\begin{tabular}{| c| c|c|}\hlinewd{1pt}
  Pressure [GPa] &	Topological Indices  &	Topological Class\\\hline
0.0 & Trivial Insulator  &  -- \\ \hline
0.5 & Trivial Insulator  &  -- \\ \hline
1.0 & Trivial Insulator & -- \\ \hline
2.0 & z$_{2w,1}$=0  z$_{2w,2}$=0  z$_{2w,3}$=0  z$_{4}$=3  &  1 \\ \hline
3.0 & z$_{2w,1}$=0  z$_{2w,2}$=0  z$_{2w,3}$=0  z$_{4}$=3  &  1 \\\hline
4.0 & z$_{2w,1}$=0  z$_{2w,2}$=0  z$_{2w,3}$=0   z$_{4}$=3  & 1 \\\hline
8.0 & z$_{2w,1}$=0  z$_{2w,2}$=0  z$_{2w,3}$=0   z$_{4}$=3  & 1 \\\hline
10.0 & z$_{2w,1}$=0  z$_{2w,2}$=0  z$_{2w,3}$=0   z$_{4}$=3  & 1 \\\hline
\hlinewd{1pt}
\end{tabular}
\end{center}
\end{table}
\endgroup

%%%%%%%%%%%%%%%%%%%%%%%%%%%%%%%%%%%%
\subsection{Lattice Dynamics}
%%%%%%%%%%%%%%%%%%%%%%%%%%%%%%%%%%%%
%%%%%%%%%%%%%%%%%%%%%%%%%%%%%%%%%%%%
\subsubsection{Phonon Dispersion Curves}
\label{sub:pdc}
%%%%%%%%%%%%%%%%%%%%%%%%%%%%%%%%%%%%

The primitive-cell of rhombohedral $\beta$-As$_2$Te$_3$ contains five atoms: Te(1) occupying the 3a Wyckoff position, and Te(2) and As(1) both at 6c. The eigenvectors corresponding to the 3D atomic displacements of each atom will therefore total 15 modes, with the three acoustic IR-active modes formed by the irreducible representations of $\Gamma_\textrm{acoustic}$ = A$_\textrm{2u}$ + E$_\textrm{u}$ and the remaining 12 optical modes being $\Gamma_\textrm{optical}$ = 2E$_\textrm{g}$ (Raman) + 2A$_\textrm{1g}$ (Raman) + 2E$_\textrm{u}$ (IR) + 2A$_\textrm{2u}$ (IR).

\begin{figure*}[!]
\begin{center}
\includegraphics[width=15cm]{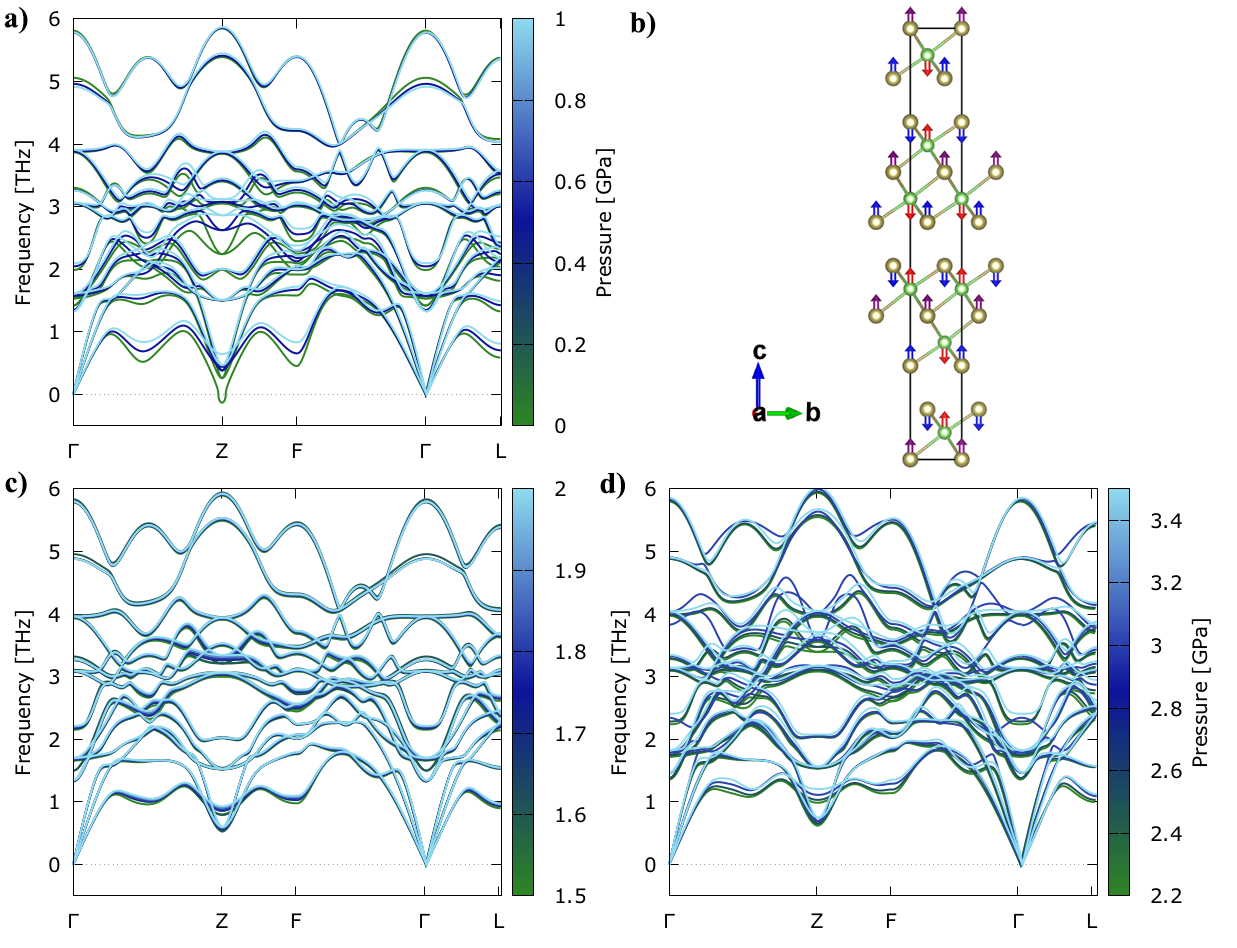}
\caption{\label{fig:phonon_disp} Phonon dispersion curves of $\beta$-As$_2$Te$_3$ for pressure values between a) 0 and 1 GPa, with b) representation of the eigenvectors associated to the negative mode at \textbf{Z}-point represented in the unit-cell (red arrows correspond to motions of the As atom, the blue and purple arrows correspond to motions of the Te(1) and Te(2) atoms, respectively). Phonon dispersion curves of $\beta$-As$_2$Te$_3$ for pressure values between c) 1.5 and 2 GPa and d) 2.2 and 3.5 GPa.} 
\end{center}
\end{figure*}

\begin{figure*}[h]
\begin{center}
\includegraphics[width=14cm]{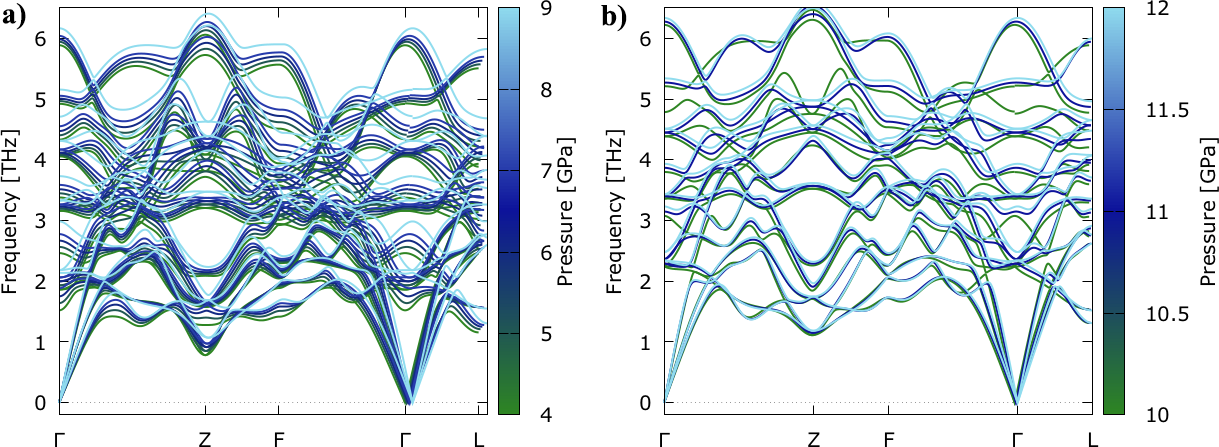}
\caption{\label{fig:phonon_supp} Phonon dispersion curves of $\beta$-As$_2$Te$_3$ for pressure values between a) 4 and 9 GPa and b) 10 and 12 GPa.} 
\end{center}
\end{figure*}

According to the phonon dispersion curves calculated for different pressure values, up to 12 GPa (see Fig. \ref{fig:phonon_disp} and Fig. \ref{fig:phonon_supp} in Suppl. Mat.), the tetradymite structure is dynamically stable up to 12 GPa. We must however note that at 0 GPa a small localized instability is observed at the high symmetry \textbf{Z}-point, which corresponds to the out-of-phase displacements between the As inter-layer atoms (Fig. \ref{fig:phonon_disp}.b; breathing mode). Such an instability persists for increasing convergence parameters (supercell size, \textbf{k}-point mesh), therefore concluding that the observed negative mode is not a numerical feature of the employed methodology. A similar imaginary mode at the same high-symmetry point has been reported by Vaney \textit{et al.}\cite{C6RA01770C} at 0 K and 0 GPa for $\beta$-As$_2$Te$_3$. By increasing the pressure up until 0.5 GPa, the negative frequency observed at the \textbf{Z}-point hardens, allowing the system to become dynamically stable. Moreover, the low-frequency phonon branches along the $\Gamma$-\textbf{Z}-\textbf{F} segment are relatively soft for low pressure values when compared with those of the $\alpha$-As$_2$Te$_3$ phase.\cite{doi:10.1021/acs.jpcc.6b06049}
By increasing the pressure the phonon branches located along these mentioned segments show a considerable increase in frequency and the abrupt gradient variations (kinks with abrupt drop of frequencies) tend to fade away at high pressure, namely along $\Gamma$-\textbf{Z} and the \textbf{F}-point.

Since the tetradymite structure is stable at room conditions, as demonstrated by several experimental studies performed for this compound,\cite{SSSR.10.1431.1974,JSolidStateChem.74.277.1988, doi:10.1063/1.4950947, doi:10.1002/aelm.201400008,C8CP00431E, InorChem.54.9936.2015,C6RA01770C} we infer that the instability observed at 0 GPa, and located at the \textbf{Z}-point, appears since we do not consider the anharmonic effects for the lattice-dynamics calculations. %Note that the imaginary mode at 0 GPa corresponds to a mode of A-type symmetry. Usually this mode corresponds to an longitudinal acoustic (LA) mode. In layered materials, such as $\beta$-As$_2$Te$_3$, instabilities are usually related to the transversal acoustic (TA) phonons that have much smaller frequencies than LA phonons. 
By including the anharmonic contributions, the rhombohedral structure of As$_2$Te$_3$ at 0 GPa would probably tend to stabilize, thus suggesting that these effects can be very important for this compound, namely at the low pressure regime. Future calculations by including the anharmonic effects should be considered to fully characterize the vibrational properties of $\beta$-As$_2$Te$_3$, however this perspective is out of the scope of the present study.

It has been observed in other related works, that the soft phonon modes can be potentially induced by a Kohn anomaly (frequency kink/dip in the phonon dispersion at certain high symmetry points) which are associated with the topological singularities of Dirac nodes, in analogy to similar effects found for graphene\cite{PhysRevLett.97.266407} and other Weyl semi-metals.\cite{PhysRevResearch.1.033101, arXiv:1906.00539} A discontinuity of the derivative of the dispersion relation is observed, when an abrupt change in the electronic screening of lattice vibrations by conduction electrons occurs (anomalies of the dielectric tensors).

The Kohn anomaly is one of the most important anomalies also observed for \textit{d}-block transition metals.\cite{PhysRevB.100.075145} The lattice vibrations are partly screened by virtual electronic excitations on the Fermi surface. This screening can change rapidly at certain wave-vector points of the BZ so the phonon energy can vary abruptly with the wave-vector.
Consequently, it usually shows a singularity or sharp dip in the phonon dispersions and a maximum in the phonon linewidth (inverse of the phonon lifetimes and detailed in Subsec. \ref{thermalconductivity}). It is believed that the Kohn anomaly can efficiently affect the superconductivity of some conventional superconductors, the lattice-dynamical instability, and the formation of spin density-waves in elemental metals.\cite{PhysRevB.100.075145}

\begin{figure*}
\begin{center}
\includegraphics[width=16cm]{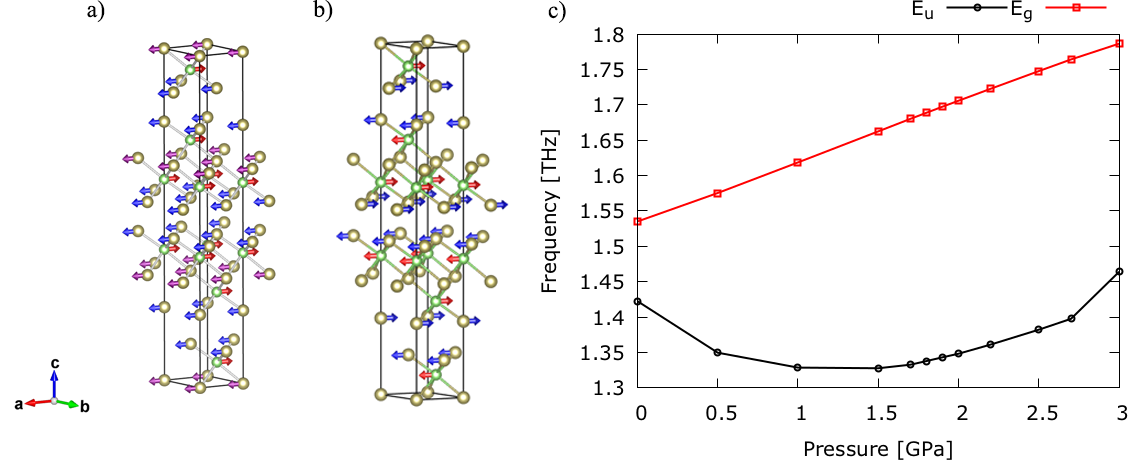}
\caption{\label{fig:soft_mode} Eigenvectors corresponding to the a) E$_u$ IR-active and b) E$_g$ Raman-active soft modes at the \textbf{$\Gamma$}-point represented in the unit-cell of $\beta$-As$_2$Te$_3$; c) and the behaviour of the two soft modes as a function of pressure.} 
\end{center}
\end{figure*}

It must be stressed that the observed Kohn anomalies occur at \textbf{q}=2k$_F$, where k$_F$ is the wave-vector where the Dirac cones appear. For the case of $\beta$-As$_2$Te$_3$ at $\sim$2 GPa, the Dirac cones appear at the $\Gamma$ point (k$_F$=0), therefore the Kohn anomalies are only expected to occur at the zone-center. In fact, the lowest optical mode at the $\Gamma$-point, E$_u$ (IR-active; Fig \ref{fig:soft_mode}.a) shows a minimum frequency around 1.33 THz between 1.0 and 1.7 GPa (Fig. \ref{fig:soft_mode}.c), while it increases at other pressure values. 
The second observed soft-mode, E$_g$ (Raman-active; Fig. \ref{fig:soft_mode}.b), presents the lowest frequency at 0 GPa, at 1.54 THz, increasing with ongoing pressure up to 1.79 THz at 3.0 GPa. 

The presence of these optical soft modes, namely the E$_u$ mode, present similarities to those found for $\alpha$-As$_2$Te$_3$ between 0 and 12 GPa.\cite{doi:10.1021/acs.jpcc.6b06049} These low frequency modes can be seen as an evidence for the appearance of a 3D TDS for $\beta$-As$_2$Te$_3$ at relatively low pressures, similarly as to what occurs for black phosphorous.\cite{PhysRevB.93.195434}  It is also noteworthy of mentioning that, and in agreement with our calculations, the Dirac cones are also formed at the $\Gamma$-point for the rhombohedral Sb$_2$Se$_3$ system, for pressures close to 3 GPa.\cite{PhysRevB.89.035101}

We emphasize here that the present calculations can only capture the Kohn anomaly due to static electronic screening of lattice vibrations, while a full treatment including dynamic screening effect will require the calculation of dynamic electron-phonon coupling, which is out of the scope of the present work.

%%%%%%%%%%%%%%%%%%%%%%%%%%%%%%%%%%%%
\subsubsection{Phonon Partial Density of States}
%%%%%%%%%%%%%%%%%%%%%%%%%%%%%%%%%%%%

\begin{figure*}[!]
\begin{center}
\includegraphics[width=14cm]{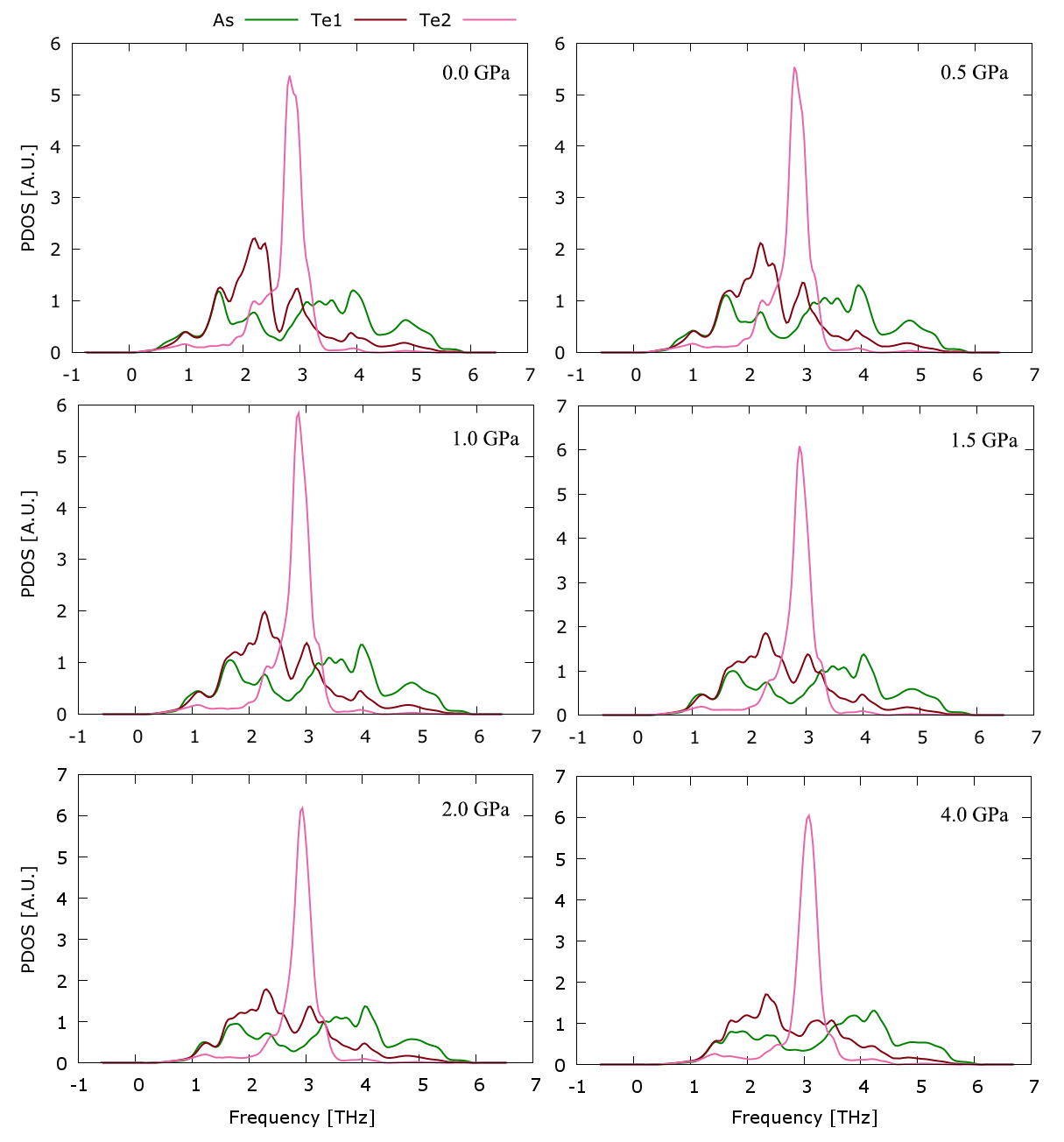}
\caption{\label{fig:pdos} Phonon partial density of states of $\beta$-As$_2$Te$_3$ for pressure values between 0 and 4 GPa.} 
\end{center}
\end{figure*}

We have computed the phonon density of states (PDoS) where the atomic contributions of the three inequivalent sites are evidenced (Fig \ref{fig:pdos}), namely Te(1) (3a site) and Te(2) and As(1) (both at 6c site). We show the PDoS as a function of pressure up to 4 GPa.  We note that the phonon dispersion curves of the $\beta$ phase (at 0 GPa) exhibits very low frequency modes, which do not exceed 6.0 THz, as to what occurs for the monoclinic phase of As$_2$Te$_3$ ($\beta$').\cite{C6RA01770C}

We observe that the Te(2) phonon states are quite localized around 3 THz, with large density of states that would correspond to interactions with the six neighboring As atoms. At 0.5 GPa, a small peak shoulder is observed around 2 THz which tends to delocalize towards lower frequencies for increasing pressures. At around 4 GPa respective states start to localize between 1 and 2 GPa. 

For lower frequencies (below 2 THz) we observe low frequency densities, namely dominated by the As and Te(1) elements. These would correspond to the E$_u$ and E$_g$ soft-modes described in Sec. \ref{sub:pdc} and related to the Kohn anomalies. The low densities observed for Te(2) at these low frequency intervals would be related to the displacements corresponding to the E$_u$ mode as observed in Fig \ref{fig:soft_mode}.a).

%%%%%%%%%%%%%%%%%%%%%%%%%%%%%%%%%%%%
\subsubsection{Lattice Thermal-Conductivity}
\label{thermalconductivity}
%%%%%%%%%%%%%%%%%%%%%%%%%%%%%%%%%%%%

\begin{figure*}
\begin{center}
\includegraphics[width=18cm]{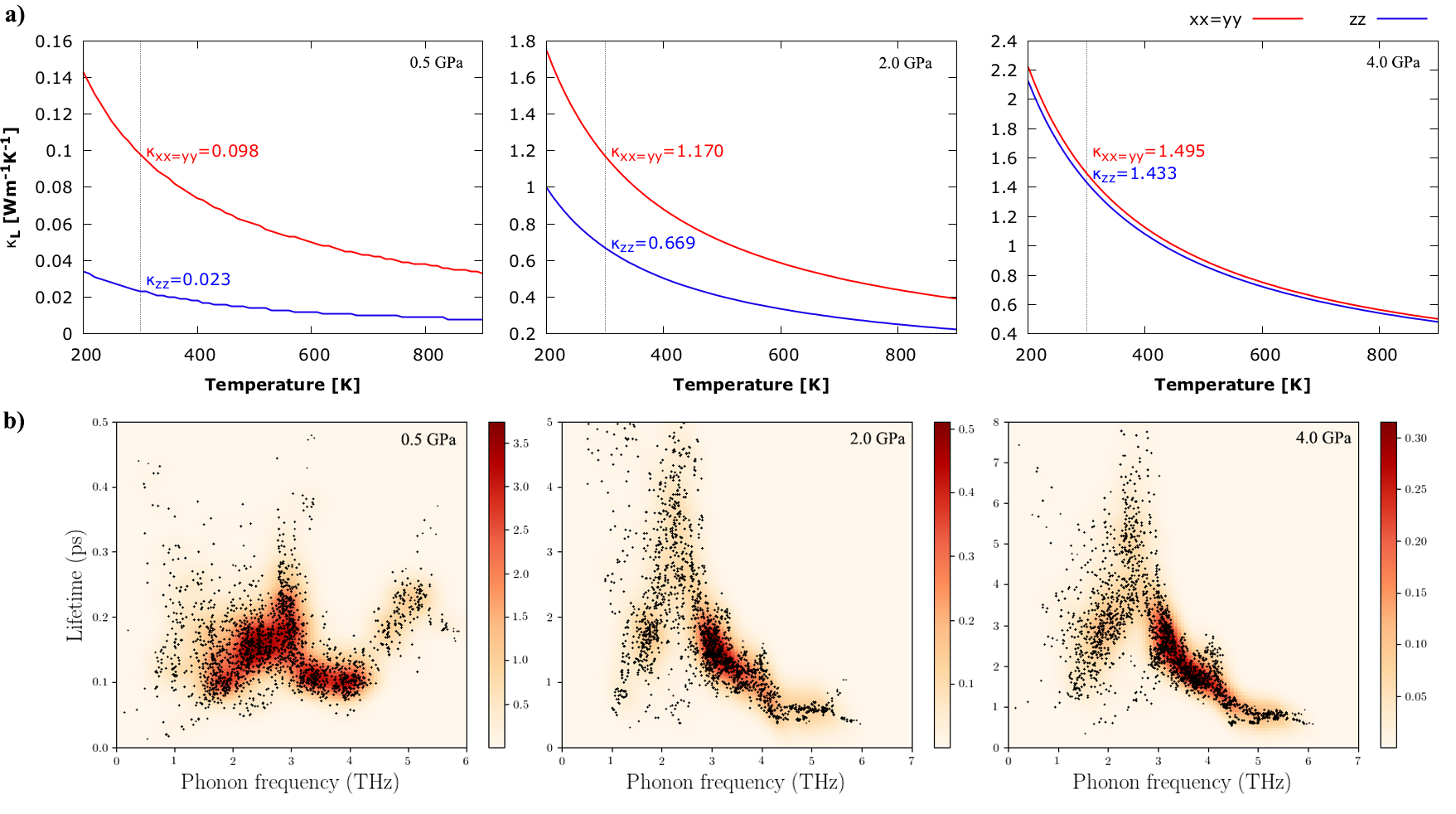}
\caption{\label{fig:kappa} a) Lattice thermal-conductivity of $\beta$-As$_2$Te$_3$ at 0.5 GPa (left), 2.0 GPa (middle) and 4.0 GPa (right). The vertical dotted line defines the T=300 K region with respective $\kappa_\textrm{L}$ value. b) Calculated phonon lifetimes of $\beta$-As$_2$Te$_3$ at 300K for 0.5 GPa (left), 2.0 GPa (middle) and 4.0 GPa (right). The color shades represent the phonon density, where darker shades refer to higher phonon densities.} 
\end{center}
\end{figure*}

Based on the third-order interatomic force constants, we have calculated the lattice thermal-conductivity ($\kappa_\textrm{L}$) for $\beta$-As$_2$Te$_3$ for pressure values of 0.5, 2.0 and 4.0 GPa (Fig. \ref{fig:kappa}.a). Due to the dynamical instability observed at 0 GPa, we have computed the $\kappa_\textrm{L}$ at 0.5 GPa. The calculation at 2 GPa, has been performed to probe $\kappa_\textrm{L}$ at the observed topological invariance and at 4 GPa which is after the QTPT. Fig. \ref{fig:kappa} shows the temperature-dependant lattice thermal-conductivity along the $x$-axis (similar values for $y$) and $z$-axis of the unit-cell (hexagonal representation) at 300 K. It is quite surprising to observe that the $\kappa_\textrm{L}$ of $\beta$-As$_2$Te$_3$ is quite low, mainly at the low-pressure limit. For 0.5 GPa, the room temperature $\kappa_\textrm{L}$ is 0.098 Wm$^{-1}$K$^{-1}$ along the two crystallographic $x$- and $y$-axis (in-plane); along the layered $z$-axis (out-of-plane), $\kappa_\textrm{L}$ lowers to 0.023 Wm$^{-1}$K$^{-1}$. At 2 GPa this value increases to 1.170 Wm$^{-1}$K$^{-1}$ and 0.669 Wm$^{-1}$K$^{-1}$, for the in-plane and out-of-plane directions, respectively. At 4 GPa we observe that the value of $\kappa_\textrm{L}$ along the out-of-plane axis increases remarkably to 1.433 Wm$^{-1}$K$^{-1}$, closing up to the in-plane value of 1.495 Wm$^{-1}$K$^{-1}$. This feature means that pressure will compress the layers to such a distance such that the van-der-Waal effects will be surpassed by the new rearrangement of the bonding environment, which will be similar to that of the in-plane.

Moreover for $\beta$-As$_2$Te$_3$ it has experimentally been observed that there is a temperature dependence of the thermal-conductivity, for which $\kappa_\textrm{L}$ decreases monotonically with increasing temperature, following roughly a T$^{–1}$ power law. Such a feature suggests that Umklapp scattering events dominate the thermal transport in this temperature range.\cite{doi:10.1002/aelm.201400008,doi:10.1063/1.4950947} The $\kappa_\textrm{L}$ values at room temperature, measured in the parallel direction were found to be low, of the order of 0.5 Wm$^{-1}$K$^{-1}$, however higher than the value obtained in the present calculations at 0.5 GPa.  
It must be stressed that the thermal-conductivity values at 0.5 GPa obtained for $\beta$-As$_2$Te$_3$ are lower than those measured for other chalcogenide systems such as SnSe, which is currently one of the most efficient thermoelectric materials. For the low-symmetry \textit{Pnma} phase (300K) of SnSe, $\kappa_\textrm{L}$ was found to be 1.43, 0.52 and 1.88 Wm$^{-1}$K$^{-1}$ along the $x$-, $y$- and $z$-axis, respectively. For the high-symmetry phase, \textit{Cmcm}, it has been observed that the isotropic average of the lattice thermal-conductivity at 800 K decreases to 0.33 Wm$^{-1}$K$^{-1}$,\cite{PhysRevLett.117.075502}. These latter values are higher than the present results obtained for $\beta$-As$_2$Te$_3$ at 0.5 GPa, however lower when compared to our results at 2 GPa. 

At 2 GPa, when $\beta$-As$_2$Te$_3$ becomes a 3D TDS, the calculated values of the thermal-conductivity at 300 K along the out-of-plane axis of $\beta$-As$_2$Te$_3$, (Fig. \ref{fig:kappa}), are around the same range as those predicted for the 3D TDS ZrTe$_5$ at 0 GPa, which along the similar axis is 0.33 Wm$^{-1}$K$^{-1}$.\cite{doi:10.1021/acsami.8b12504}
Moreover we note that the room temperature results at are close to those predicted for the 3D TDS Na$_2$MgPb (1.77 and 0.81 Wm$^{-1}$K$^{-1}$ along the in-plane and out-of-plane axis, respectively); for which the low lattice thermal-conductivity is mainly due to the short phonon lifetimes of the system.\cite{PhysRevB.99.024310} Our results are however slightly larger than those recently calculated for the 3D TDS Bi ($\sim$ 0.1 Wm$^{-1}$K$^{-1}$) and claimed to be among the lowest value ever found for crystalline materials.\cite{arxiv:2001.10124v1}

Moreover, regarding the prototypical Cd$_3$As$_2$ it has been shown that the existence of soft optical phonon modes affects the lattice thermal-conductivity (which ranges from 0.3 up to 0.7 Wm$^{-1}$K$^{-1}$ at 300 K).\cite{PhysRevResearch.1.033101} The low-frequency optical phonon modes increases the available phase space of the phonon-phonon scattering of heat-carrying acoustic phonons. Consequently this effect will cause the low lattice thermal-conductivity values for Cd$_3$As$_2$; which also occurs for other known thermoelectric materials, i.e PbTe,\cite{PhysRevLett.112.175501} and SnSe.\cite{NaturePhys.11.1063.2015} Furthermore, it has been shown that the interplay between the phonon-phonon Umklapp scattering rates and the soft optical phonon frequency explains the unusual non-monotonic temperature dependence of the lattice thermal conductivity of Cd$_3$As$_2$.\cite{PhysRevResearch.1.033101} 
Such a feature of the low-frequency optical phonon modes, is also observed in the phonon dispersion curves of the present calculations for $\beta$-As$_2$Te$_3$ (Fig. \ref{fig:phonon_disp}), mostly at the low pressure regime.

It has been claimed that Kohn anomalies are indicative of structural instabilities. Indeed our calculations for $\beta$-As$_2$Te$_3$ show that the system is dynamically unstable at room pressure (Fig. \ref{fig:phonon_disp}.a). The low-energy phonons mainly at the \textbf{Z}-point clearly show that $\beta$-As$_2$Te$_3$ is indeed metastable above 0.5 GPa. Such a metastable state can lead to lower values of the lattice thermal-conductivity than that obtained when $\beta$-As$_2$Te$_3$ is a 3D TDS (around 2 GPa). Therefore, this work may pave the way to search for structural instabilities which occur in 3D TDS by tuning different parameters such as temperature, pressure, chemical composition, in order to search for the best conditions to obtain the lowest thermal conductivity compatible with the highest ZT value. 

Moreover, we have to state that heavy atoms have low vibrational frequencies which consequently result in a low lattice thermal-conductivity.\cite{C5TC02344K} These tend to also exhibit large spin–orbit coupling necessary for certain nontrivial topological materials. In addition, TIs often have a small electronic band-gap as they lie in the vicinity of a strain dependent QTPT which aids in tuning the intrinsic carrier concentration to optimize the thermal-conductivity. 

Materials with resonant bonding (lead chalcogenides, SnTe, Bi$_2$Te$_3$, Bi and Sb), currently known as metavalent bonding,\cite{doi:10.1002/adfm.201904862,doi:10.1002/adma.201904316,doi:10.1002/adma.201803777,doi:10.1002/adma.201806280} commonly evidence long-ranged interactions.\cite{PMID:24770354} Long-ranged interactions may be another cause for optical phonon softening, strong anharmonic scattering and large phase space for three-phonon scattering processes, for which such features can explain the reason why rocksalt IV–VI compounds have much lower thermal-conductivity than the zincblende III–V compounds.\cite{PMID:24770354,PhysRevMaterials.2.120302} In fact, it has been observed through first-principles calculations, that long-ranged interactions are significant in IV–VI materials owing to the strong resonant bonding or hybridization between different electronic configurations).\cite{PMID:24770354}

In order to further understand the intrinsic lattice thermal-conductivity as a function of applied pressure, we have also calculated the frequency-dependent phonon lifetimes at 0.5 GPa, 2 GPa and 4 GPa (Fig. \ref{fig:kappa}.b), all at 300 K. In principle the anharmonicity of a material will be inversely related to the phonon lifetime, and larger anharmonicity will result in lower lattice thermal-conductivity. 

We find that the frequency-dependent phonon lifetimes of $\beta$-As$_2$Te$_3$ at 300K and 0.5 GPa is very short, roughly located below 0.5 ps; much lower than those found for SnSe (from 0 to 30 ps).\cite{PhysRevB.92.115202} Also our values of the phonon lifetimes are lower than those found for ZrTe$_5$ and comparable to those of Na$_3$Bi.\cite{arxiv:2001.10124v1}
In the mid-frequency region a larger density of phonons are located between 1.5 THz and $\sim$ 4.2 THz with maximal value at 0.2 ps. The small lifetimes of the phonons indicate a strong scattering rate, which is the main reason for the low lattice thermal-conductivity at this pressure range.\cite{PhysRevB.99.024310} At 2 GPa (Fig. \ref{fig:kappa}.b), the lifetimes increase to higher values reaching 5.0 ps. The larger density of phonons are more localized and concentrated around 2.5 and 4.0 THz, with maximal value located below 2.0 ps. The densities scattered between 1 and 2 THz are related to the soft-modes which are the main source of low lattice thermal-conductivity due to the Kohn anomalies, and these evidence larger lifetimes. At 4 GPa the larger density is located between 2.5 and 4 THz, similarly as to what occurs for 2 GPa, however with larger lifetimes (roughly below 4 ps). The larger lifetimes are observed at 8 ps and are mainly the contribution of phonons ranging between 2 and 3 THz.
Consequently, our calculations of phonon lifetimes allow us to explain the nature of the phonons which cause the low lattice thermal-conductivity of $\beta$-As$_2$Te$_3$ and the increase of the thermal-conductivity as pressure increases.

%%%%%%%%%%%%%%%%%%%%%%%%%%%%%%%%%%%%
\section{Conclusions}
%%%%%%%%%%%%%%%%%%%%%%%%%%%%%%%%%%%%

We have theoretically investigated the electronic and phonon dispersion curves and the lattice thermal-conductivity of $\beta$-As$_2$Te$_3$ as a function of pressure. 

Our \textit{QS}GW electronic electronic band structure reveals that at room conditions $\beta$-As$_2$Te$_3$ is a trivial semiconductor, which band-gap closes as pressure increases, until it undergoes a QTPT close to 2 GPa. At this pressure range, the material becomes a 3D TDS with well-defined 3D massless charge carriers at the $\Gamma$ point. Above this pressure the gap reopens and the material becomes a strong topological insulator. Finally, above 4 GPa, the gap tends to decrease again so that an insulator-metal transition occurs above 6 GPa with topological features persisting up until 12 GPa.

Moreover, we have investigated the lattice dynamics and lattice thermal-conductivity of $\beta$-As$_2$Te$_3$ at selected pressures. We have identified the existence of two soft optical phonons (E$_u$ and E$_g$) at the zone-center that are related to the Kohn anomaly associated with the Dirac nodes close to 2 GPa. Unlike other 3D TDS systems, however similarly to what is observed for other compounds with $R\bar{3}m$ structure, $\beta$-As$_2$Te$_3$ does not show Kohn anomalies at the zone-boundaries; these are only observed at the Brillouin-zone center. This feature will inhibit the appearance of any type of Peierls distortion,\cite{doi:10.1002/adma.201908302} as it has recently been questioned for related pure tetradymite-like materials.

Based on the observation of the soft modes, we explained that the low lattice thermal-conductivity is caused by the optical soft-modes which enhance the phonon-phonon scatterings, in a similar manner as to what occurs for the prototypical 3D TDS Cd$_3$As$_2$, ZrTe$_5$ and Na$_3$Bi. 
By comparing the lattice thermal-conductivity at three different pressure points related to the QTPT: before, at the transition point and after the QTPT; we find that mainly at low pressure, i.e. 0.5 GPa, the lattice thermal-conductivity is very low when compared to 3D TDSs, such as ZrTe$_5$ and Na$_2$MgPb, and of prototypical thermoelectric materials, i.e SnSe, isostructural Bi$_2$Te$_3$. Therefore we conclude that close to room pressure, it is possible to engineer and enhance the thermoelectric properties of $\beta$-As$_2$Te$_3$.

Moreover, we predict the possibility of finding materials with very low lattice thermal-conductivity values among materials that can be driven by different perturbations (temperature, pressure, composition, etc.) and which are close to a structural instability derived from the Kohn anomalies originated from linear dispersion effects (Dirac or Weyl cones).

%%%%%%%%%%%%%%%%%%%%%%%%%%%%%%%%%%%%
\section{Theoretical Framework}
%%%%%%%%%%%%%%%%%%%%%%%%%%%%%%%%%%%%

Density functional theory (DFT)\cite{hohenberg-pr-136-1964,PhysRev.140.A1133.1965} calculations have been preformed within the framework implemented in the Vienna Ab-initio Simulation Package (VASP) code.\cite{PhysRevB.47.558.1993,PhysRevB.54.11169.1996} The semi-local generalized-gradient approximation functional with the Perdew-Burke-Ernzerhof parametrization revised for solids (PBEsol)\cite{perdew-prl-100-2008,perdew-prl-102-2009} was employed for the structural relaxations, lattice dynamics and thermal-conductivity calculations. Projector augmented-wave (PAW) \cite{PhysRevB.59.1758.1999,PhysRevB.50.17953.1994} pseudopotentials were used to treat semi-core electronic states, with the As[4s$^2$4p$^3$] and Te[5s$^2$5p$^4$] electrons being included in the valence shell.  The starting point for our calculations was a full structural relaxation of the $R\bar{3}m$ phase, performed with a plane-wave kinetic-energy cut-off of 800 eV. The electronic BZ was sampled with a $\Gamma$-centred Monkhorst-Pack mesh \cite{monkhorst-prb-13-1976} and defined with 14$\times$14$\times$14 subdivisions.  

The theoretical background regarding the harmonic lattice-dynamics calculations is presented in Refs. \cite{PhysRevB.81.174301.2010,PhysRevB.91.144107.2015}, and therefore it will not be detailed in the present work. Lattice-dynamics calculations were performed using the supercell finite-displacement method implemented in the Phonopy software package,\cite{togo-prb-78-2008} with VASP used as the 2nd order force-constant calculator.\cite{chaput-prb-84-2001} Calculations of the phonon supercell size were carried out on 2$\times$2$\times$2 expansions of the primitive-cell. The phonon frequencies were sampled on an interpolated 50$\times$50$\times$50 \textbf{q}-point mesh (tetrahedron method) when evaluating the phonon DoS and vibrational internal energy and entropy. 

Lattice thermal-conductivity and phonon lifetimes were calculated by employing the Phono3py code,\cite{PhysRevB.91.094306.2015} and VASP is used as the calculator for the third-order (anharmonic) interactomic force-constants. 
A 2$\times$2$\times$2 supercell is used, with a \textbf{q}-mesh of 50$\times$50$\times$50 with the tetrahedron method to perform the integration for the phonon lifetime calculation. The phonon lifetimes were computed with the single-mode relaxation-time approximation, to solve the Boltzmann transport equations.

Calculations to obtain the electronic band-structure for different pressure values were performed using the Questaal (formerly LMSuite) package.\cite{QUESTAAL} Questaal is an all-electron implementation of density-functional theory and the quasiparticle self-consistent form (\textit{QS}GW).\cite{PhysRevLett.96.226402.2006,PhysRevB.76.165106.2007} The basis-sets applied to expand the wavefunctions are defined with a combination of smoothed Hankel functions and augmented plane-waves, known as the Plus Muffin-Tin (PMT) basis-sets.\cite{PhysRevLett.96.226402.2006,PhysRevB.76.165106.2007} Spin-orbit coupling (SoC) was included for all electronic structure calculations.  

For the \textit{QS}GW calculations the BZ was sampled using the tetrahedron method \cite{PhysRevB.50.17953.1994} with a sampling mesh of 6$\times$6$\times$6 subdivisions. The plane-wave cut-off for the interstitial charge density (GMAX) was defined with a 6 Ry cut-off radius. For the \textit{QS}GW calculation the G-vector cut-offs for the interstitial part of the eigenfunctions and the Coulomb interaction matrix were set to 6.0 and 5.4 Ry, respectively.

\clearpage
\section{Appendix}

%%%%%%%%%%%%%%%%%%%%%%%%%%%%%%%%%%%%%%%%%%%%
\subsection{Phonon Dispersion Curves for Pressures above 4 GPa}
\label{sub:phonon}
%%%%%%%%%%%%%%%%%%%%%%%%%%%%%%%%%%%%%%%%%%%%

\clearpage
%
%%%%%%%%%%%%%%%%%%%%%%%%%%%%%%%%%%%%%%%%%%%%
\subsection{Topological Data}
%%%%%%%%%%%%%%%%%%%%%%%%%%%%%%%%%%%%%%%%%%%%

\begin{table*}[h]
\begin{center}
\caption{ \label{table:topo_list} Topological character of translation equivalent subgroups for the system at 2.0, 4.0 and 8.0 GPa, where we present the number and symbol of the space group, the transformation matrix, the possibility of forming linear combinations of the elementary band representations (EBR) below the Fermi level (the system is a topological insulator if one cannot form EBRs). Minimal subgroups are highlighted in the table.}
\begin{tabular}{| c| c|c|c|}\hlinewd{1pt}
  Symmetry Group &	Transformation matrix& 	EBR & Topological Indices  \\\hline
 1 \textit{P1} & 2/3,1/3,1/3,1/3,2/3,-1/3,-1/3,1/3,1/3,0,0,0 & yes & -- \\
 \rowcolor{gray!25}{2 \textit{P-1}} & 2/3,1/3,1/3,1/3,2/3,-1/3,-1/3,1/3,1/3,0,0,0 & no & z$_{2w,1}$=0  z$_{2w,2}$=0  z$_{2w,3}$=0  z$_{4}$=3 \\
 146 \textit{R3} & 1,0,0,0,1,0,0,0,1,0,0,0 & yes & -- \\
 148 \textit{R-3} & 1,0,0,0,1,0,0,0,1,0,0,0 & no & z$_{2w,1}$=0  z$_{2w,2}$=0  z$_{2w,3}$=0  z$_{4}$=3\\ \hline
\hlinewd{1pt}
\end{tabular}
\end{center}
\end{table*}

\clearpage

%%%%%%%%%%%%%%%%%%%%%%%%%%%%%%%%%%%%
\section{Acknowledgements}
%%%%%%%%%%%%%%%%%%%%%%%%%%%%%%%%%%%%

This research was supported by the European Union Horizon 2020 research and innovation programme under Marie Sklodowska-Curie grant agreement No. 785789-COMEX and project NORTE-01-0145-FEDER-022096, Network of Extreme Conditions Laboratories (NECL), financed by FCT and co-financed by NORTE 2020, through the programme Portugal 2020 and FEDER. Authors also thank the financial support of the Generalitat Valenciana under Project PROMETEO 2018/123-EFIMAT and of the Agencia Española de Investigación under Projects MAT2016-75586-C4-2/4-P, FIS2017-2017-83295-P, PID2019-106383GB-C42, as well as the MALTA Consolider Team research network under project RED2018-102612-T. Additionally, authors acknowledge the computer resources at MareNostrum with technical support provided by the Barcelona Supercomputing Center (QCM-2019-1-0032/37).

\clearpage

%\bibliographystyle{unsrt}
%%%\bibliographystyle{abbrv} 
%\bibliography{biblo}

\end{document}